\documentclass{INTERSPEECH2023}
\usepackage{caption}
\usepackage{graphicx}
\usepackage{float} 
\usepackage{subcaption}
\usepackage{url}
\interspeechcameraready

\title{EmoMix: Emotion Mixing via Diffusion Models for Emotional Speech Synthesis}  
\name{Haobin Tang$^{1,2\dagger}$, Xulong Zhang$^{1\dagger}$, Jianzong Wang$^{1\ast}$, Ning Cheng$^{1}$, Jing Xiao$^{1}$
\thanks{$^\dagger$Equal Contribution}
\thanks{$^\ast$Corresponding author: Jianzong Wang, jzwang@188.com.}}
\address{$^{1}$Ping An Technology (Shenzhen) Co., Ltd., Shenzhen, China\\$^{2}$University of Science and Technology of China, Hefei, China}

\begin{document}

\maketitle
 
\begin{abstract}
There has been significant progress in emotional Text-To-Speech (TTS) synthesis technology in recent years. However, existing methods primarily focus on the synthesis of a limited number of emotion types and have achieved unsatisfactory performance in intensity control. To address these limitations, we propose EmoMix, which can generate emotional speech with specified intensity or a mixture of emotions. Specifically, EmoMix is a controllable emotional TTS model based on a diffusion probabilistic model and a pre-trained speech emotion recognition (SER) model used to extract emotion embedding. Mixed emotion synthesis is achieved by combining the noises predicted by diffusion model conditioned on different emotions during only one sampling process at the run-time. We further apply the \emph{Neutral} and specific primary emotion mixed in varying degrees to control intensity. Experimental results validate the effectiveness of EmoMix for synthesizing mixed emotion and intensity control.
\end{abstract}    

\section{Introduction}
As the seq2seq modeling architecture continues to evolve rapidly, reference-based style transfer is an effective method for emotional speech synthesis. Recent researches follow the "say it like this" principle and explored various approaches, such as Global Style Tokens (GST)~\cite{wang2018style}, as well as its updates~\cite{ma2019neural}. Phoneme and Segmental-level prosody embedding are used in Generspeech~\cite{huanggenerspeech} to capture a wide range of emotional variations across different scales. To model more disentangled prosody embedding some studies~\cite{liu2021expressive,li22h_interspeech} use an intermediate embedding extracted from a certain layer of Speech emotion recognition (SER) or automatic speech recognition (ASR) model as deep emotional features. Promising results are achieved in the aspect of emotion expressiveness but the task of achieving intensity controllable emotional TTS remains challenging. Various methods have been proposed to manipulate internal emotion representations, such as scaling~\cite{choi2021sequence}, interpolation~\cite{um2020emotional}, or distance-based quantization~\cite{im2022emoq}. The introduction of relative attributes~\cite{parikh2011relative} has been proposed in emotional TTS~\cite{lei2022msemotts,zhu2019controlling} and emotion voice conversion (EVC) ~\cite{zhou2022emotion} to develop a more properly defined and calculated emotion intensity values. 

However, previous methods predominantly concentrate on synthesizing a limited number of emotion types. Few frameworks have explored the correlations and interactions between different emotions. 
Humans are capable of experiencing approximately 34,000 distinct emotions and even multiple emotional states simultaneously~\cite{plutchik2001nature,braniecka2014mixed}. To address this issue, Plutchik~\cite{plutchik2013theories} proposed a set of eight primary emotions, namely sadness, disgust, joy, fear, anger, anticipation, surprise, and trust. All other emotions can be considered as derived forms or a mix of these primary emotions. For example, the mixture of \emph{Happy} and \emph{Surprise} can be regarded as \emph{Excitement}. Building a large mixed emotion data set will bring a lot of cost in practical application. Reference-based unseen style transfer may be a solution for mixed emotion synthesis, but it is also difficult to obtain a suitable mixed emotion reference audio set. 
MixedEmotion~\cite{zhou2022speech} study the modeling of mixed emotions in TTS for the first time. 
MixedEmotion obtains intensity values from relative attributes rank~\cite{parikh2011relative} to weight the emotion embedding. The desired mixture of multiple emotions is achieved by manually defining an emotion attribute vector. However, this brings a noticeable degradation in quality. 
EmoDiff~\cite{guo2022emodiff} uses a soft-label guidance technique based on the classifier guidance ~\cite{DBLP:conf/nips/DhariwalN21} in denoising diffusion probabilistic models (DDPM)~\cite{ho2020denoising,songscore} to synthesize controllable and mixed emotion. But this classifier must be trained on noisy audio and would be inefficient for high dimensional and unseen primary emotion conditioning.

This study introduces EmoMix, a framework for transferring emotional speech styles that employs a diffusion probabilistic model and a pre-trained speech emotion recognition (SER) model.
We use emotion embedding extracted by SER model as an extra condition to enable the reverse process of diffusion model for primary emotion generation. Moreover,  to overcome intensity control and mixed emotion issues, we avoid directly modeling the mixed emotion by introducing mix methods~\cite{liew2022magicmix,kim2022diffusionclip} from image semantic mixing tasks. Specifically, the noises predicted by DDPM model based on different emotional conditions are combined to synthesize mixed emotions through only one sampling process at the run-time. The intensity is further controlled via \emph{Neutral} and specific emotion combinations.
\begin{figure*}[!t]   
\centering
\includegraphics[width=0.8\linewidth]{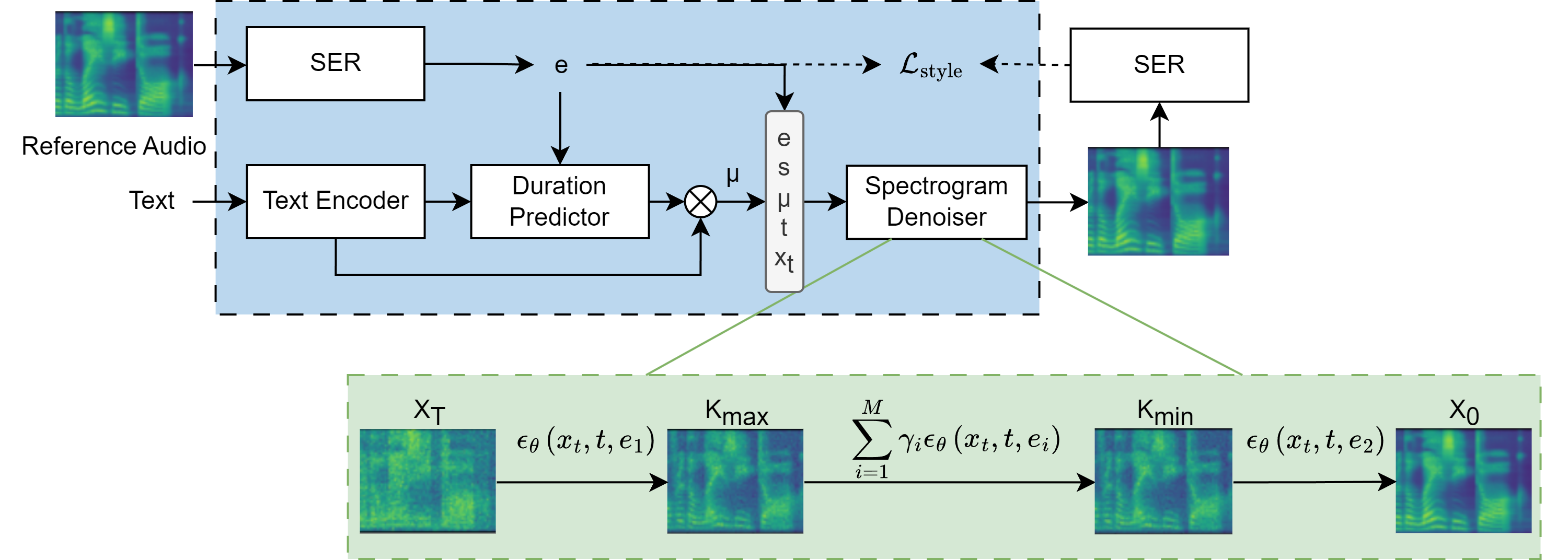}
\caption{The overview architecture for EmoMix and the green part represents the extended sampling process of mixed emotion synthesis at run-time. The dotted arrows are only used during training. The SER mdoel is pre-trained and their parameters are frozen.}
\label{Fig1}
\end{figure*}
In summary, the primary advantages of EmoMix are:
\begin{enumerate}
    \item We use high dimensional emotional embedding extracted from reference audio by pre-trained SER model for seen and unseen primary emotion conditioning instead of the discrete emotion label used by classifier guidance.
    \item EmoMix achieves mixed emotion generation in terms of a noise combination during only one sampling process at run-time, needless for directly modeling mixed emotion. Furthermore, 
    EmoMix is can generate speech with a specified primary emotional intensity through the process of mixing \emph{Neutral} and target primary emotions in varying proportions.
    \item EmoMix can mix and control emotions without posing harm to synthesized voice quality.
    The generated audio samples exhibit high levels of naturalness and quality. 
\end{enumerate}

\section{EmoMix}
\subsection{Preliminary on Score-based Diffusion Model}
Denoising Diffusion Probablistic Models (DDPM) ~\cite{ho2020denoising} has recently achieved great performance in image~\cite{liew2022magicmix,kim2022diffusionclip} and audio~\cite{popov2021grad,huang2022prodiff} generation tasks. We propose EmoMix, a novel method for emotion mixing in speech synthesis based on the design of GradTTS~\cite{popov2021grad}, which applies stochastic differential equation (SDE)~\cite{songscore} formulation to TTS. GradTTS defines the diffusion process, which can convert any data distribution $X_0$ to standard normal distribution $X_T$ as follows:
\begin{equation}
\label{eq1}
    d X_{t}=-\frac{1}{2} X_{t} \beta_{t} d t+\sqrt{\beta_{t}} d W_{t}, \quad t \in[0, T]
\end{equation}
where $\beta_{t}$ is the pre-defined noise schedule and $W_{t}$ is the Wiener process. As Anderson proves SDE can formulate the reverse process, which follows the diffusion process’s reverse trajectory:
\begin{equation}
\label{eq2}
    d X_{t}=\left(-\frac{1}{2} X_{t}-\nabla_{X_{t}} \log p_{t}\left(X_{t}\right)\right) \beta_{t} d t+\sqrt{\beta_{t}} d \widetilde{W_{t}}
\end{equation}
GradTTS uses the following equation, which is a discretized version of the reverse $\mathrm{SDE}$, during sampling to generate data $X_{0}$ from standard Gaussian noise $X_{T}$
\begin{equation}
\label{eq3}
    X_{t-\frac{1}{N}}=X_{t}+\frac{\beta_{t}}{N}\left(\frac{1}{2} X_{t}+
    \nabla_{X_{t}}\log p_{t}(X_{t}\right))+\sqrt{\frac{\beta_{t}}{N}} z_{t},
\end{equation}
where $N$ is the number of steps in the discretized reverse process, and $z_{t}$ is the standard Gaussian noise. As GradTTS set $T$ to 1, the size of one step is $\frac{1}{N}$, and $t \in\left\{\frac{1}{N}, \frac{2}{N}, \ldots, 1\right\}$.

Given data $X_{0}$, $X_{t}$ is sampled from the distribution derived from Eq.~\eqref{eq1} to estimate the score $\nabla_{X_{t}}\log p_{t}(X_{t})$ as follows:
\begin{equation}
\label{eq4}
    X_{t} \mid X_{0} \sim \mathcal{N}\left(\rho\left(X_{0}, t\right), \lambda(t)\right)
\end{equation}
where $\rho\left(X_{0}, t\right)=\mathrm{e}^{-\frac{1}{2} \int_{0}^{t} \beta_{s} d s} X_{0}$, and $\lambda(t)=I-$ $\mathrm{e}^{-\int_{0}^{t} \beta_{s} d s}$. Eq.(3) derives the score, $\nabla_{X_{t}} \log p_{t}\left(X_{t} \mid X_{0}\right)=-\lambda(t)^{-1} \epsilon_{t}$, where $\epsilon_{t}$ is the standard Gaussian noise used to sample $X_{t}$ given $X_{0}$. To estimate the score $\epsilon_{\theta}(X_{t}, t, \mu, s, e)$ is trained for $\forall t \in[0, T]$. Where $\mu$ is phoneme-dependent Gaussian mean conditioned on speaker $s$ and emotion $e$. 
We always need text and speaker as condition and we focus on emotion throughout the paper. Therefore $\mu$ and $s$ in $\epsilon_{\theta}(X_{t}, t, \mu, s, e)$ are omitted, and the network is expressed as a simplified notation $\epsilon_{\theta}(X_{t}, t, e)$. The following loss is used:
\begin{equation}
\label{eq5}
\mathcal{L}_{diff} = \mathbb{E}_{\boldsymbol{x}_0, t, \boldsymbol{e},\epsilon_t}[\lambda_t \mathbb{||} \boldsymbol{\epsilon}_\theta (\boldsymbol{x}_t, t, \boldsymbol{e}) +{\lambda(t)}^{-1} \epsilon_t  \mathbb{||}_2^2]
\end{equation}




\subsection{Emotion Conditioning with SER}
\label{section:2.2}
Previous work~\cite{guo2022emodiff} only considers single-speaker emotional TTS and uses the gradient of log-probability of classifier to guide the reverse process into a limited number of discrete emotion categories. Inspired by DeepEST~\cite{zhou2021seen}, we leverage a pre-trained SER model that produces a continuous emotion embedding $e$ from a reference utterance that exemplifies the target emotional prosody to condition the model on the desired emotion. The SER architecture is the same as that in~\cite{chen20183}. Initially, 3-D CNN layer takes the mel-spectrum and its derivatives as input and extracts a latent representation that encodes the emotional content while filtering out the irrelevant factors. The BLSTM and attention layer then generate an emotion embedding $e$ that represents the utterance-level feature for emotion classification. For speaker conditioning, we use wav2vec 2.0 model~\cite{baevski2020wav2vec} in Generspeech~\cite{huanggenerspeech} to capture the speaker acoustic condition $s$. 

As illustrated in Figure~\ref{Fig1}, EmoMix is based on GradTTS, except that the predicted duration is conditioned on emotion and speaker. Hidden representations $\mu$ reflects the linguistic contents from input text, emotion embedding $e$ and speaker embedding $s$. Consequently, the spectrogram denoiser iteratively refines $\mu$ into mel-spectrograms with target primary emotion and speaker in reference audio. 

We use another SER along with the denoiser to further minimize the emotion style gap between the reference speech and the synthesized one. Style loss using gram matrix is a widely used technique in computer vision~\cite{johnson2016perceptual} and has been recently applied to measure mel-spectrogram~\cite{ma2019neural} and emotion embedding~\cite{li2021controllable}. To maintain emotion prosody of reference audio in synthesized speech, we suggest a style reconstruction loss:
\begin{equation}
\label{eq6}
    \mathcal{L}_{\text {style}}=\sum_j\mathbb{||}G_j(\hat{m}) - G_j(m) \mathbb{||}_F^2
\end{equation}
where $G_j(x)$ is the gram matrix of the j-th layer feature map of the 3-D CNN in SER model for the input x. $m$ and $\hat{m}$ denote reference mel-spectrogram and synthesized mel-spectrogram respectively. This style reconstruction loss enforces synthesized speech to have a similar style to that of the reference audio. The final training objective is:
\begin{equation}
\label{eq7}
    \mathcal{L} = \mathcal{L}_{\text {dur}} + \mathcal{L}_{\text {diff}} + \mathcal{L}_{\text {prior}} + \gamma\mathcal{L}_{\text {style}}
\end{equation}
where $\mathcal{L}_{\text {dur}}$ is the $\ell_{2}$ loss of logarithmic duration, and $\mathcal{L}_{\text {diff }}$ is the diffusion loss as Eq.\eqref{eq5}. $\gamma$ is hyperparameter and empirically set to 1e-4. We also adopt prior loss $\mathcal{L}_{\text {prior}}$ in GradTTS to encourage converging. 

\subsection{Run-time Emotion Mixing}
Our goal is to synthesize speech with mixed emotions or with a single primary emotion of varying intensities at the run-time. 
As shown in the green section of Figure~\ref{Fig1}, we extend the reverse process of the trained DDPM, which was originally conditioned on a single primary emotion in section 2.2, to synthesize mixed emotions. Mix methods ~\cite{liew2022magicmix,kim2022diffusionclip} was first used in computer vision to solve semantic mixing task which aims to modify the content in a certain part of a given object in image while preserving its layout semantics. It is believed that the sampling procedure of DDPMs first crafts coarse features and details appear last~\cite{ho2020denoising}. 

EmoMix enables emotion mixing of two different emotions by replacing the condition vector after sampling step $K_{max}$ during inference with the purpose of overwriting the base emotion details with the mixed-in emotion. We average the emotion embedding of a set of audio samples that have same primary emotion to avoid the instability of single reference audio~\cite{cai2021emotion}. We first synthesize coarse base emotion prosody by denoising from Gaussian noise given the base emotion condition $e_1$ (e.g., \emph{Happy}) to the intermediate steps $K_{max}$, followed by denoising on condition of the mixed-in emotion $e_2$ (e.g., \emph{Surprise}) from $K_{min}$ to obtain a mixture of emotion (e.g., \emph{Excitement}). 
From timestep $K_{max}$ to $K_{min}$, we apply the noise combine approach~\cite{kim2022diffusionclip} to better preserve elements of the given base emotion and prevent it from being too easily overwritten by the mixed-in emotion. We combine the noises predicted from multiple emotion conditions to synthesize multiple emotion styles through only one sampling process. Specifically, we combine the multiple noises according to the following rule:
\begin{equation}
\label{eq8}
     \boldsymbol\epsilon_{\theta}  \left(\boldsymbol{x}_{t}, t, e_{mix}\right)
     =\sum_{i=1}^{M} \gamma_{i} \boldsymbol\epsilon_{\theta}\left(\boldsymbol{x}_{t}, t, e_i\right)
\end{equation}
where $\gamma_i$ is the weight of each condition $e_i$ satisfying $\sum_{i=1}^{M} \gamma_{i}=1$, which can be used for controlling the degree of each emotion. M is the number of mixed-in emotion categories and is set to 2 for dual emotion mixture. Therefore, EmoMix allows for flexible mixing of emotions in various combinations, without having to train new models with target mixed emotion condition that is directly modeled. According to Eq.\eqref{eq8}, this sampling process can be interpreted as increasing the joint probability of the following conditional distributions:
\begin{equation}
 \sum_{i=1}^{M} \gamma_{i} \boldsymbol\epsilon_{\theta}\left(\boldsymbol{x}_{t}, t, e_i\right) \propto-\nabla_{\boldsymbol{x}_{t}} \log \prod_{i=1}^{M} p\left(\boldsymbol{x}_{t} \mid e_{\mathrm{tar}, i}\right)^{\gamma_{i}}   
\end{equation}
where $e_{\mathrm{tar}, \mathrm{i}}$ is the specified target emotion condition.

Inspired by the interpolation~\cite{um2020emotional} technique, we intuitively apply the \emph{Neutral} and primary emotion mixture for  emotion intensity control. By mixing the noise from the \emph{Neutral} condition and the specified primary emotion with varying $\gamma$, we can perform interpolation between the \emph{Neutral} and the target emotion smoothly to control the primary emotion intensity.

\section{Experiment}
\subsection{Experiment Setup}
The SER model is trained on IEMOCAP~\cite{busso2008iemocap} which contains 10k utterances in total to obtain the emotional features $e$. A subset of IEMOCAP with five emotion types namely \emph{Sad}, \emph{Surprise}, \emph{Happy}, \emph{Neutral} and \emph{Angry}. We use the English part of the ESD dataset~\cite{zhou2021seen} which contains 10 speaker with the same five emotions used in IEMOCAP to perform experiments. We adopt the same data partition as the ESD dataset and the \emph{Angry} is left as unseen primary emotion for the experiments.

The noise estimation network $\epsilon_\theta$ consists of the U-Net and linear attention modules that are identical to those in GradTTS. EmoMix is trained with 32 batch size and Adam optimizer at $10^{-4}$ learning rate for 1M steps. We provide the alignment of speech and text extracted by Montreal Forced Aligner (MFA)~\cite{mcauliffe2017montreal} for training duration predictor. We employed HifiGAN~\cite{kong2020hifi} as the vocoder for subsequent experiments.

\subsection{Evaluation of Emotional Speech}
\label{section:3.2}
We evaluate the quality and similarity of synthesized audio samples of EmoMix against following baseline models:
\begin{enumerate}   
  \item GT and GT (voc.): ground truth audio and wav generated from ground truth mel-spectrogram using HiFi-GAN.
  \item MixedEmotion~\cite{zhou2022speech}: A relative attribute ranking-based model which pre-compute intensity values for training and can be manually controlled at run time.
  \item EmoDiff~\cite{guo2022emodiff}: EmoDiff is a controllable emotional TTS which can synthesize diverse emotional speech based on DDPM and guided by soft-label.
\end{enumerate}
Note that in this experiment and ablation study, samples from EmoMix are conditioned on single primary emotion during reverse process and mix method is not used. 

For subjective evaluation, assessors are asked to rate 20 speech per emotion for naturalness (mean opinion score; MOS) and similarity to the target emotion (similarity mean opinion score; SMOS) measures on a scale from 1 to 5 with 1 point intervals. For objective evaluation, we use mel cepstral distortion (MCD)~\cite{kubichek1993mel} to evaluate the spectrum similarity between the reference mel-spectrum and generated features. Table~\ref{table1} presents that the vocoder has negligible effect on speech quality. EmoMix surpasses baselines in SMOS with a significant margin while having comparable MOS and reaching the lowest MCD results. We observe that EmoMix still achieves satisfactory results when synthesizing unseen emotions (\emph{Angry}). The results indicate the superiority of EmoMix over the baselines.
\begin{table}[htbp]
	\centering
	\caption{ Results for single primary emotion synthesis.}
	\scalebox{0.9} {
		\begin{tabular}{lccc}
            \hline
            \textbf{Model} & \textbf{MOS $\uparrow$}  & \textbf{SMOS $\uparrow$}  & \textbf{MCD $\downarrow$} \\
            \hline
            GT & $4.47 \pm 0.08$ & $4.43 \pm 0.08$ & $-$ \\
            GT (voc.) & $4.40 \pm 0.10$ & $4.38 \pm 0.08$& $2.87$ \\
            \hline
            MixedEmotion~\cite{zhou2022speech} & $3.61 \pm 0.08$ & $3.85 \pm 0.09$& $6.17$ \\
            EmoDiff~\cite{guo2022emodiff} & $4.08 \pm 0.12$ & $3.87 \pm 0.10$& $5.76$ \\  
            \hline
            EmoMix (seen) & $4.10 \pm 0.10$ & $4.02 \pm 0.08$& $5.29$ \\
            EmoMix (unseen) & $3.92 \pm 0.10$ & $3.82 \pm 0.12$& $5.65$ \\
            \hline
            \end{tabular}
	}
	\label{table1}
\end{table}

\begin{figure*}[htbp]
	\centering
	\begin{subfigure}{0.23\linewidth}
		\centering
		\includegraphics[width=1\linewidth]{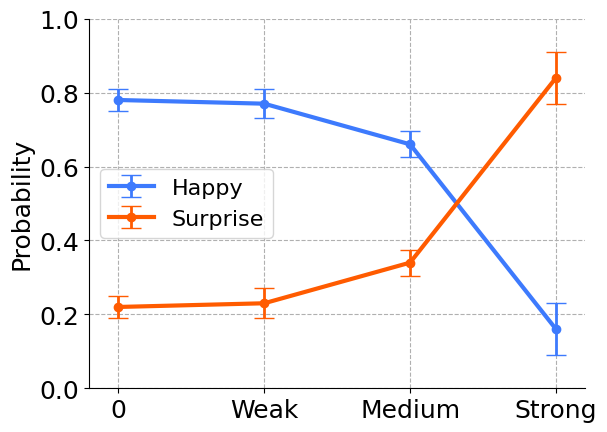}
		\caption{Happy-Surprise (Excitement)}
		\label{fig2-1}
	\end{subfigure}
	\centering
	\begin{subfigure}{0.23\linewidth}
		\centering
		\includegraphics[width=1\linewidth]{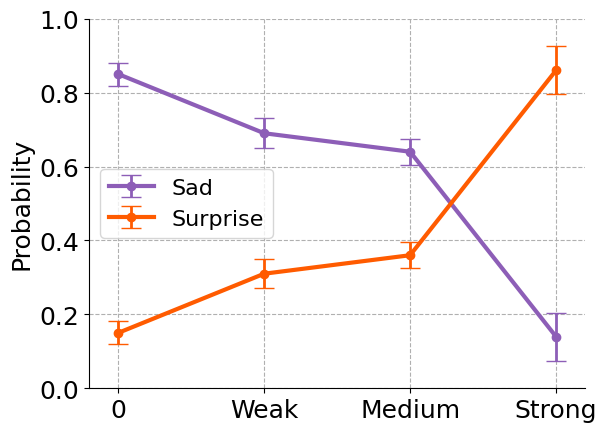}
		\caption{Happy-Sad (Disappointment)}
		\label{fig2-2}
	\end{subfigure}
        \centering
	\begin{subfigure}{0.23\linewidth}
		\centering
		\includegraphics[width=1\linewidth]{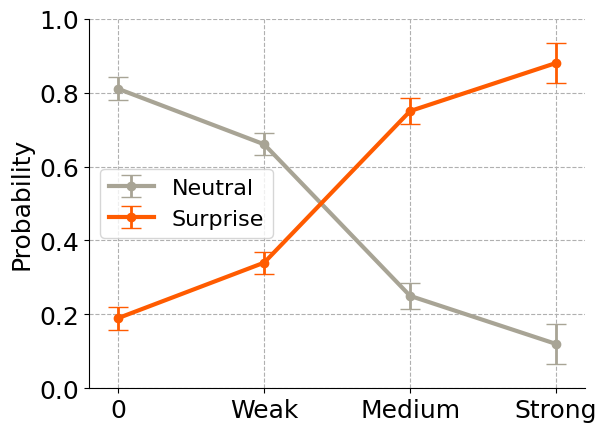}
		\caption{Neutral-Surprise}
		\label{fig2-3}
	\end{subfigure}
	\centering
	\begin{subfigure}{0.23\linewidth}
		\centering
		\includegraphics[width=1\linewidth]{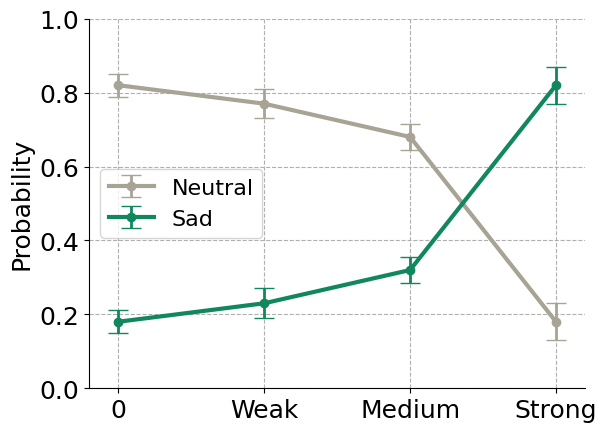}
		\caption{Neutral-Sad}
		\label{fig2-4}
	\end{subfigure}
	\caption{Classification probabilities obtained from the pre-trained SER model. Each point indicates the mean probability value of 20 utterances with mixed two emotions, while the probability of other emotions is negligible and thus omitted.}
 \vspace{-1em}
	\label{Fig2}
\end{figure*}

\subsection{Ablation Study}
To assess the contribution of the techniques used in EmoMix, including utilizing SER and the style reconstruction loss, we perform ablation studies and present the results in Table~\ref{table2}. To compare the quality and expressiveness of the generated speeches, comparative mean opinion score (CMOS) and comparative similarity mean opinion score (CSMOS) are used. We use GradTTS (w/ emo label) denotes GradTTS model conditioned on discrete emotion labels instead of the emotion features extracted by SER. The drop of quality and similarity scores indicate the importance of modeling emotion style representation in continuous space. We can also observe that the absence of style reconstruction loss lead to obvious decline in CSMOS. This indicates that the style loss can force EmoMix to synthesize target emotion in the reference audio.
\begin{table}[ht]
	\centering
	\caption{CMOS and CSMOS comparison for ablation study.}
        \begin{tabular}{lcc}
            \hline\hline
            \textbf{Model}& \textbf{CMOS} & \textbf{CSMOS} \\
            \hline 
            EmoMix & /& /   \\
            \hline
            GradTTS (w/ emo label) & -0.05 & -0.16   \\
            EmoMix (w/o $\mathcal{L}_{\text {style}}$) & -0.01 & -0.09   \\
            \hline\hline
        \end{tabular}

	\label{table2}
 \vspace{-1.5em}
\end{table}


\begin{table}[!b]
\vspace{-1em}
	\centering
	\caption{CMOS comparison for mixed emotion.}
	\scalebox{0.9} {
        \begin{tabular}{lcc}
            \hline\hline
            \textbf{Mix Intensity}& \textbf{EmoMix} & \textbf{MixedEmotion}~\cite{zhou2022speech} \\
            \hline 
            w/o mix & / & /     \\
            Weak & -0.11 & -0.23    \\
            Medium & -0.07 & -0.22  \\
            Strong & -0.06 & -0.19  \\
            \hline\hline
        \end{tabular}
	}
	\label{table3}
\end{table}

\subsection{Evaluation of Mixed Emotion}
We fix $K_{max}$ and $K_{min}$ to be $0.6T$ and $0.2T$ resepectively and vary only $\gamma$ to control mixed-in emotion intensity where the larger $\gamma$ indicates a stronger intensity of the mixed-in emotion. $T$ is set to 10 in this part. 
It is challenging for assessors to perceive the subtle emotion differences under continuous $\gamma$ settings. Therefore, we set the $\gamma \in [0.1, 0.3], [0.4, 0.6]$ and $[0.7, 0.8]$ to represent the weak, medium and strong emotion intensity level respectively. We set $\gamma$ no lager than 0.8 to prevent base emotion to be completely overwritten.

Two different mixed emotions namely \emph{Excitement} and \emph{Disappointment} are evaluated. We use \emph{Happy} and \emph{Sad} as base emotion respectively and mix \emph{Surprise} in. These mixed emotions have been studied in emotion theory~\cite{plutchik2013theories}.
We follow MixedEmotion~\cite{zhou2022speech} to employ the classification probabilities obtained from the softmax layer of another SER that is pre-trained on ESD dataset to analyze the performance of mixed emotions. We present the evaluation of different mixed-in intensity of each mixed emotion in Figure~\ref{Fig2}(a), (b). 
 We notice that the probability of \emph{Surprise} consistently rises and that of \emph{Happy} and \emph{Sad} decrease when gradually increase the percentage of noise corresponding to \emph{Surprise}.  These observations indicate that by varying the percentage of mixed-in emotion noise, we can easily synthesize and control the desired mixed emotion which can be recognized by a SER. Moreover, we present CMOS result of 10 types of mixed emotion in different mixed-in intensity in Table~\ref{table3}. The mix method of EmoMix brings slightly decreases in speech quality while that of MixedEmotion result in obvious quality degradation. These results indicate that EmoMix can generate and control mixed emotional speech without compromising the quality effectively.

\begin{figure}[!t]   
\centering
\includegraphics[width=0.8\linewidth]{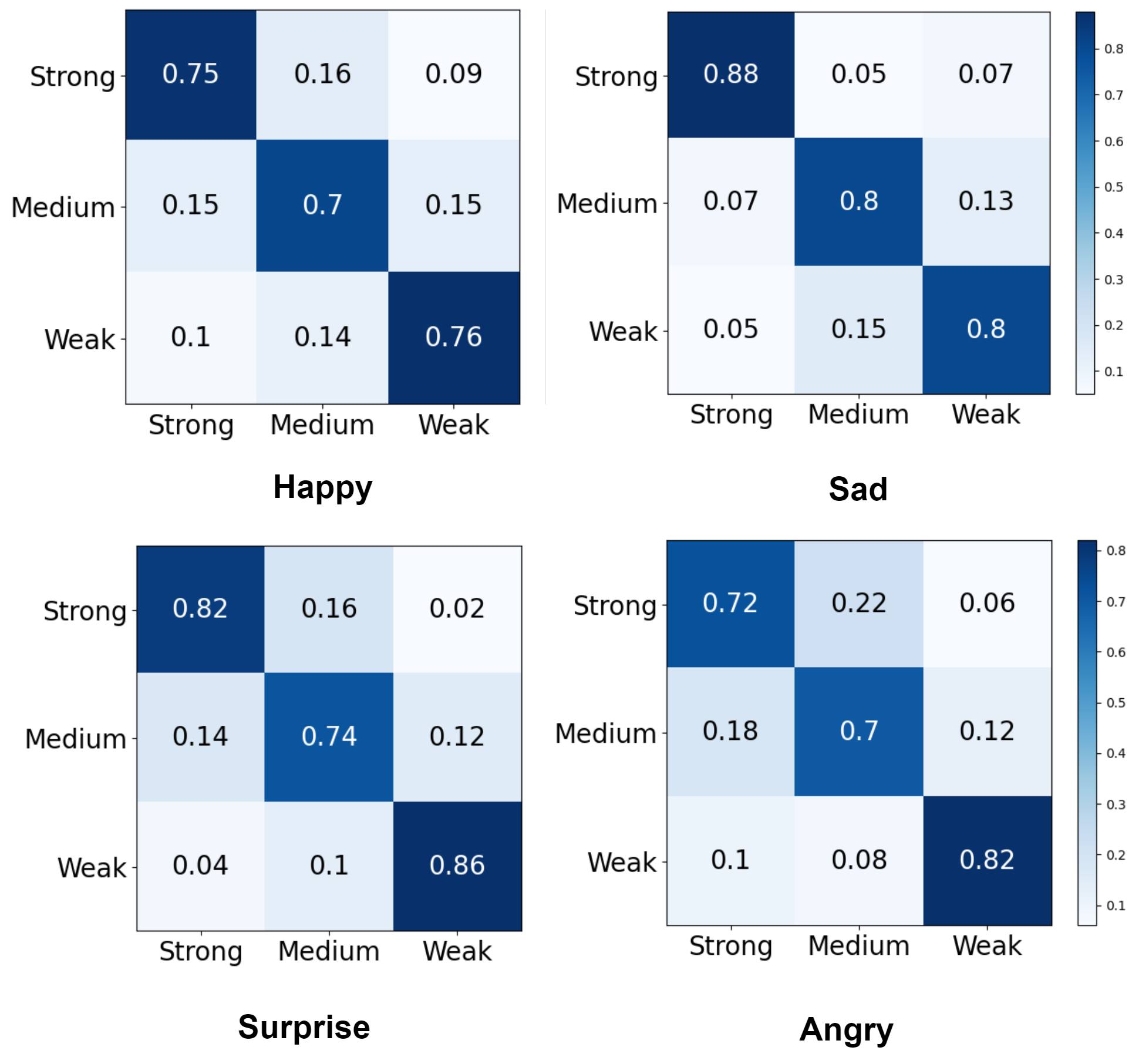}
\caption{ Confusion matrices of synthesized mixed emotion. The vertical and horizontal axes of each subplot indicate the  ground truth and perceived emotion intensity, respectively.}
\label{Fig3}
\vspace{-1em}
\end{figure}

Then \emph{Neutral} is further introduced as base emotion for primary emotion intensity control. An increase of mixed-in emotion similar to that seen in Figure 2(a), (b) appears in Figure 2(c), (d). This demonstrates that \emph{Neutral} and primary emotion mixing will lead to diverse primary emotions. To further evaluate the intensity of primary emotion a human perceptual experiment was conducted. Figure~\ref{Fig3} illustrates the confusion matrix for each emotion category obtained from the evaluation.
The results clearly demonstrate that the generated samples accurately reflect the desired intensity, which confirms the effectiveness of EmoMix in controlling primary emotion intensity.

\section{Conclusion}
We propose a controllable emotional TTS framework, EmoMix, to further explore mixed emotion synthesis and intensity control. We avoid explicitly modeling the mixed emotion by introducing mix methods. By manually combining the noise at runtime, EmoMix could produce different emotional mixtures. Evaluations demonstrate the ability to generate speech with various mixed emotions.

\section{Acknowledgement}
Supported by the Key Research and Development Program of Guangdong Province (grant No. 2021B0101400003) and Corresponding author is Jianzong Wang (jzwang@188.com).

\bibliographystyle{IEEEtran}
\bibliography{EmoMix}

\begin{thebibliography}{10}
\providecommand{\url}[1]{#1}
\csname url@samestyle\endcsname
\providecommand{\newblock}{\relax}
\providecommand{\bibinfo}[2]{#2}
\providecommand{\BIBentrySTDinterwordspacing}{\spaceskip=0pt\relax}
\providecommand{\BIBentryALTinterwordstretchfactor}{4}
\providecommand{\BIBentryALTinterwordspacing}{\spaceskip=\fontdimen2\font plus
\BIBentryALTinterwordstretchfactor\fontdimen3\font minus
  \fontdimen4\font\relax}
\providecommand{\BIBforeignlanguage}[2]{{%
\expandafter\ifx\csname l@#1\endcsname\relax
\typeout{** WARNING: IEEEtran.bst: No hyphenation pattern has been}%
\typeout{** loaded for the language `#1'. Using the pattern for}%
\typeout{** the default language instead.}%
\else
\language=\csname l@#1\endcsname
\fi
#2}}
\providecommand{\BIBdecl}{\relax}
\BIBdecl

\bibitem{wang2018style}
Y.~Wang, D.~Stanton, Y.~Zhang, R.-S. Ryan, E.~Battenberg, J.~Shor, Y.~Xiao,
  Y.~Jia, F.~Ren, and R.~A. Saurous, ``Style tokens: Unsupervised style
  modeling, control and transfer in end-to-end speech synthesis,'' in
  \emph{International Conference on Machine Learning}.\hskip 1em plus 0.5em
  minus 0.4em\relax PMLR, 2018, pp. 5180--5189.

\bibitem{ma2019neural}
S.~Ma, D.~Mcduff, and Y.~Song, ``Neural tts stylization with adversarial and
  collaborative games,'' in \emph{International Conference on Learning
  Representations}, 2019.

\bibitem{huanggenerspeech}
R.~Huang, Y.~Ren, J.~Liu, C.~Cui, and Z.~Zhao, ``Generspeech: Towards style
  transfer for generalizable out-of-domain text-to-speech,'' in \emph{Advances
  in Neural Information Processing Systems}, vol.~35, 2022, pp.
  10\,970--10\,983.

\bibitem{liu2021expressive}
R.~Liu, B.~Sisman, G.~Gao, and H.~Li, ``Expressive tts training with frame and
  style reconstruction loss,'' \emph{IEEE/ACM Transactions on Audio, Speech,
  and Language Processing}, vol.~29, pp. 1806--1818, 2021.

\bibitem{li22h_interspeech}
T.~Li, X.~Wang, Q.~Xie, Z.~Wang, M.~Jiang, and L.~Xie, ``{Cross-speaker Emotion
  Transfer Based On Prosody Compensation for End-to-End Speech Synthesis},'' in
  \emph{Proc. Interspeech 2022}, 2022, pp. 5498--5502.

\bibitem{choi2021sequence}
H.~Choi and M.~Hahn, ``Sequence-to-sequence emotional voice conversion with
  strength control,'' \emph{IEEE Access}, vol.~9, pp. 42\,674--42\,687, 2021.

\bibitem{um2020emotional}
S.-Y. Um, S.~Oh, K.~Byun, I.~Jang, C.~Ahn, and H.-G. Kang, ``Emotional speech
  synthesis with rich and granularized control,'' in \emph{ICASSP 2020-2020
  IEEE International Conference on Acoustics, Speech and Signal Processing
  (ICASSP)}.\hskip 1em plus 0.5em minus 0.4em\relax IEEE, 2020, pp. 7254--7258.

\bibitem{im2022emoq}
C.-B. Im, S.-H. Lee, S.-B. Kim, and S.-W. Lee, ``Emoq-tts: Emotion intensity
  quantization for fine-grained controllable emotional text-to-speech,'' in
  \emph{ICASSP 2022-2022 IEEE International Conference on Acoustics, Speech and
  Signal Processing (ICASSP)}.\hskip 1em plus 0.5em minus 0.4em\relax IEEE,
  2022, pp. 6317--6321.

\bibitem{parikh2011relative}
D.~Parikh and K.~Grauman, ``Relative attributes,'' in \emph{2011 International
  Conference on Computer Vision}.\hskip 1em plus 0.5em minus 0.4em\relax IEEE,
  2011, pp. 503--510.

\bibitem{lei2022msemotts}
Y.~Lei, S.~Yang, X.~Wang, and L.~Xie, ``Msemotts: Multi-scale emotion transfer,
  prediction, and control for emotional speech synthesis,'' \emph{IEEE/ACM
  Transactions on Audio, Speech, and Language Processing}, vol.~30, pp.
  853--864, 2022.

\bibitem{zhu2019controlling}
X.~Zhu, S.~Yang, G.~Yang, and L.~Xie, ``Controlling emotion strength with
  relative attribute for end-to-end speech synthesis,'' in \emph{2019 IEEE
  Automatic Speech Recognition and Understanding Workshop (ASRU)}.\hskip 1em
  plus 0.5em minus 0.4em\relax IEEE, 2019, pp. 192--199.

\bibitem{zhou2022emotion}
K.~Zhou, B.~Sisman, R.~Rana, B.~W. Schuller, and H.~Li, ``Emotion intensity and
  its control for emotional voice conversion,'' \emph{IEEE Transactions on
  Affective Computing}, no.~01, pp. 1--1, 2022.

\bibitem{plutchik2001nature}
R.~Plutchik, ``The nature of emotions: Human emotions have deep evolutionary
  roots, a fact that may explain their complexity and provide tools for
  clinical practice,'' \emph{American Scientist}, vol.~89, no.~4, pp. 344--350,
  2001.

\bibitem{braniecka2014mixed}
A.~Braniecka, E.~Trzebi{\'n}ska, A.~Dowgiert, and A.~Wytykowska, ``Mixed
  emotions and coping: The benefits of secondary emotions,'' \emph{PloS one},
  vol.~9, no.~8, p. e103940, 2014.

\bibitem{plutchik2013theories}
R.~Plutchik and H.~Kellerman, \emph{Theories of emotion}.\hskip 1em plus 0.5em
  minus 0.4em\relax Academic Press, 2013, vol.~1.

\bibitem{zhou2022speech}
K.~Zhou, B.~Sisman, R.~Rana, B.~W. Schuller, and H.~Li, ``Speech synthesis with
  mixed emotions,'' \emph{IEEE Transactions on Affective Computing}, 2022.

\bibitem{guo2022emodiff}
Y.~Guo, C.~Du, X.~Chen, and K.~Yu, ``Emodiff: Intensity controllable emotional
  text-to-speech with soft-label guidance,'' in \emph{IEEE International
  Conference on Acoustics, Speech and Signal Processing (ICASSP)}, 2023, pp.
  1--5.

\bibitem{DBLP:conf/nips/DhariwalN21}
P.~Dhariwal and A.~Q. Nichol, ``Diffusion models beat gans on image
  synthesis,'' in \emph{Advances in Neural Information Processing Systems},
  vol.~34, 2021, pp. 8780--8794.

\bibitem{ho2020denoising}
J.~Ho, A.~Jain, and P.~Abbeel, ``Denoising diffusion probabilistic models,''
  \emph{Advances in Neural Information Processing Systems}, vol.~33, pp.
  6840--6851, 2020.

\bibitem{songscore}
Y.~Song, J.~Sohl-Dickstein, D.~P. Kingma, A.~Kumar, S.~Ermon, and B.~Poole,
  ``Score-based generative modeling through stochastic differential
  equations,'' in \emph{International Conference on Learning Representations},
  2021.

\bibitem{liew2022magicmix}
J.~H. Liew, H.~Yan, D.~Zhou, and J.~Feng, ``Magicmix: Semantic mixing with
  diffusion models,'' \emph{arXiv preprint arXiv:2210.16056}, 2022.

\bibitem{kim2022diffusionclip}
G.~Kim, T.~Kwon, and J.~C. Ye, ``Diffusionclip: Text-guided diffusion models
  for robust image manipulation,'' in \emph{Proceedings of the IEEE/CVF
  Conference on Computer Vision and Pattern Recognition}, 2022, pp. 2426--2435.

\bibitem{popov2021grad}
V.~Popov, I.~Vovk, V.~Gogoryan, T.~Sadekova, and M.~Kudinov, ``Grad-tts: A
  diffusion probabilistic model for text-to-speech,'' in \emph{International
  Conference on Machine Learning}.\hskip 1em plus 0.5em minus 0.4em\relax PMLR,
  2021, pp. 8599--8608.

\bibitem{huang2022prodiff}
R.~Huang, Z.~Zhao, H.~Liu, J.~Liu, C.~Cui, and Y.~Ren, ``Prodiff: Progressive
  fast diffusion model for high-quality text-to-speech,'' in \emph{Proceedings
  of the 30th ACM International Conference on Multimedia}, 2022, pp.
  2595--2605.

\bibitem{zhou2021seen}
K.~Zhou, B.~Sisman, R.~Liu, and H.~Li, ``Seen and unseen emotional style
  transfer for voice conversion with a new emotional speech dataset,'' in
  \emph{ICASSP 2021-2021 IEEE International Conference on Acoustics, Speech and
  Signal Processing (ICASSP)}.\hskip 1em plus 0.5em minus 0.4em\relax IEEE,
  2021, pp. 920--924.

\bibitem{chen20183}
M.~Chen, X.~He, J.~Yang, and H.~Zhang, ``3-d convolutional recurrent neural
  networks with attention model for speech emotion recognition,'' \emph{IEEE
  Signal Processing Letters}, vol.~25, no.~10, pp. 1440--1444, 2018.

\bibitem{baevski2020wav2vec}
A.~Baevski, Y.~Zhou, A.~Mohamed, and M.~Auli, ``wav2vec 2.0: A framework for
  self-supervised learning of speech representations,'' \emph{Advances in
  Neural Information Processing Systems}, vol.~33, pp. 12\,449--12\,460, 2020.

\bibitem{johnson2016perceptual}
J.~Johnson, A.~Alahi, and L.~Fei-Fei, ``Perceptual losses for real-time style
  transfer and super-resolution,'' in \emph{European Conference on Computer
  Vision (ECCV) Part II 14}, 2016, pp. 694--711.

\bibitem{li2021controllable}
T.~Li, S.~Yang, L.~Xue, and L.~Xie, ``Controllable emotion transfer for
  end-to-end speech synthesis,'' in \emph{2021 12th International Symposium on
  Chinese Spoken Language Processing (ISCSLP)}.\hskip 1em plus 0.5em minus
  0.4em\relax IEEE, 2021, pp. 1--5.

\bibitem{cai2021emotion}
X.~Cai, D.~Dai, Z.~Wu, X.~Li, J.~Li, and H.~Meng, ``Emotion controllable speech
  synthesis using emotion-unlabeled dataset with the assistance of cross-domain
  speech emotion recognition,'' in \emph{ICASSP 2021-2021 IEEE International
  Conference on Acoustics, Speech and Signal Processing (ICASSP)}.\hskip 1em
  plus 0.5em minus 0.4em\relax IEEE, 2021, pp. 5734--5738.

\bibitem{busso2008iemocap}
C.~Busso, M.~Bulut, C.~Lee, A.~Kazemzadeh, E.~Mower, S.~Kim, J.~N. Chang,
  S.~Lee, and S.~S. Narayanan, ``{IEMOCAP:} interactive emotional dyadic motion
  capture database,'' \emph{Lang. Resour. Evaluation}, vol.~42, no.~4, pp.
  335--359, 2008.

\bibitem{mcauliffe2017montreal}
M.~McAuliffe, M.~Socolof, S.~Mihuc, M.~Wagner, and M.~Sonderegger, ``Montreal
  forced aligner: Trainable text-speech alignment using kaldi.'' in
  \emph{Interspeech}, vol. 2017, 2017, pp. 498--502.

\bibitem{kong2020hifi}
J.~Kong, J.~Kim, and J.~Bae, ``Hifi-gan: Generative adversarial networks for
  efficient and high fidelity speech synthesis,'' in \emph{Advances in Neural
  Information Processing Systems}, vol.~33, 2020, pp. 17\,022--17\,033.

\bibitem{kubichek1993mel}
R.~Kubichek, ``Mel-cepstral distance measure for objective speech quality
  assessment,'' in \emph{Proceedings of IEEE Pacific Rim Conference on
  Communications Computers and Signal Processing}, vol.~1.\hskip 1em plus 0.5em
  minus 0.4em\relax IEEE, 1993, pp. 125--128.

\end{thebibliography}

\end{document}